\documentstyle[aps,prd]{revtex}
\begin{document}
\title{Spacelike Fluctuations of the Stress Tensor for de Sitter Vacuum}
\author{Albert Roura and Enric Verdaguer \thanks{Institut de F\'\i sica
d'Altes Energies (IFAE)}}
\address{Departament de F\'{\i}sica Fonamental,
Universitat de Barcelona, Av.~Diagonal 647,\\
08028 Barcelona, Spain}
\maketitle

\begin{abstract}
The two-point function characterizing the stress tensor fluctuations of a
massless minimally coupled field for an invariant vacuum state in de
Sitter spacetime is discussed. This two-point function is explicitly
computed for spacelike separated points which are geodesically connected.
We show that these fluctuations are as important as the
expectation value of the stress tensor itself. These quantum field
fluctuations will induce fluctuations in the geometry of de Sitter
spacetime.
This paper is a first step towards the computation of such metric
fluctuations, which may be of interest for large-scale structure formation
in cosmology. The relevance of our results in this context is
briefly discussed.
\end{abstract}

\section{INTRODUCTION}

In today's standard inflationary scenario the amplified vacuum
fluctuations
of the inflaton field provide the seeds for large-scale structure
formation 
\cite{linde90,padmanabhan93,brandenberger92/liddle93}. It is thus of
interest to understand the effect of these quantum
fluctuations on the universe dynamics and the implications they may have
on cosmological observations.

In most inflationary models (exponential inflation), the geometry of
spacetime during the period of inflation can be reasonably well described
by
de Sitter spacetime. This spacetime also seems favored by quantum
cosmology since most
accepted ``initial condition'' proposals, which are either quantum
``tunneling from nothing'' \cite{vilenkin84/86} or ``no-boundary''
condition \cite{hartle83}, both predict it.

It has been recently pointed out that fluctuations in the stress-energy
tensor of quantum fields may be important for some states in curved
spacetimes or even flat spacetimes with non-trivial topology \cite{kuo93}.
Hu and Phillips, for instance, have computed the energy density
fluctuations
of quantum states in spatially closed Friedmann-Robertson-Walker models
and have
shown that these could be as important as the energy density itself \cite
{hu97}. If so, these fluctuations may induce relevant back-reaction
effects
on the gravitational field (the spacetime geometry).

In this paper we compute the fluctuations of the stress-energy
tensor for a scalar field in de Sitter spacetime. We consider a massless
minimally coupled field in the Euclidean vacuum state, which is a de
Sitter invariant state. The motivation for considering the massless
minimal coupling case
is the fact that it mimics the behavior of gravitons in a curved
background
as well as that of the perturbations of the inflaton field in usual
inflationary models. As for the state, the reasons why we chose the
Euclidean vacuum are twofold. On the one hand, the high degree of symmetry
makes the computations simpler. On the other hand, there are at least two
serious physical motivations. First, it naturally arises in exponential
inflation models and for a massive field it has been shown to be the state
to which any other state tends asymptotically in a de Sitter background
\cite{hawking83}; and second, it is selected by the most popular boundary
conditions in quantum cosmological models \cite{halliwell89}.

Here we compute the two-point correlation for spacelike separated points
of the
stress-energy tensor and find that these fluctuations are important.
Therefore the back reaction of these fluctuations on the spacetime
geometry
could be relevant, and will be the subject of further work within the
context of stochastic semiclassical gravity and Einstein-Langevin
equation \cite{martin98}.
Related to this fact it is worth mentioning that Abramo, Brandenberger and
Mukhanov \cite{abramo97a,abramo97b} have shown that the second order
contribution
to back reaction of inflaton and gravitational perturbations during the
inflationary period can be important even below the ``self-reproduction''
scale.
Although the spirit of our approach to the back-reaction problem is
slightly
different, we believe a partial connection with their work may exist.
These issues will also be addressed in future investigations.

The plan of the paper is the following. In section \ref{sec2} we give a
very brief review of de Sitter spacetime, its properties and the solution
of the Klein-Gordon equation in such a background. In section \ref{sec3},
following the proposal made in \cite{garriga93}, we discuss the invariant
vacuum state for a massless minimally coupled field that we are going to
use. In section \ref{sec4} we deal with the fluctuations of the
stress-energy tensor of the scalar field in de Sitter background and
compute the noise kernel (i.e. the two-point correlations), which
characterizes these fluctuations, for spacelike separated points. We
discuss our results in the final section.
Throughout the paper we will use the $(+,+,+)$ sign convention of Misner,
Thorne and Wheeler's book \cite{misner71}.

\section{DE SITTER SPACETIME} \label{sec2}

In this section we give a brief summary of some useful properties and
definitions related to de Sitter spacetime. For a more detailed
exposition, the reader is referred to \cite{hawking73/mottola85,allen85}.

Four dimensional de Sitter space-time has positive constant curvature and
is thus maximally symmetric. Its ten dimensional group of isometries is
$O(4,1)$. It can be represented as a hyperboloid embedded in a five
dimensional spacetime with Minkowskian metric $\eta _{AB}$, so that 
$\eta_{AB}\xi ^A(x)\xi ^B(x)=H^{-2}$, where the Hubble constant $H$ is
related to the scalar curvature $R=12H^2$ and $\xi ^A(x)$ is the position
vector in the
five-dimensional Minkowskian embedding spacetime (with $A,B=0,\ldots ,4$)
corresponding to the point $x$ of de Sitter. One can define the biscalar 
\begin{equation}
\label{00}Z(x,y)\equiv H^2\eta _{AB}\xi ^A(x)\xi ^B(y) ,
\end{equation}
$Z>1$ for timelike separated points, $Z=1$ for points connected by null
geodesics and $Z<1$ for spacelike separated points. It is worth
emphasizing that there is no geodesic connecting two points with
$Z<-1$. For points which are geodesically connected an alternative
expression to (\ref{00}) is \cite{allen85}: 
\begin{equation}
\label{0}Z(x,y)=\cos \sqrt{\frac{R\,\sigma (x,y)}6}, 
\end{equation}
where $\sigma (x,y)\equiv \frac 12 s^2(x,y)$ with $s(x,y)$ being the
geodesic distance
between the two points. We will use the closed coordinates $(\eta ,\chi
,\theta ,\varphi )$, which cover the whole de Sitter spacetime. The line
element reads: 
\begin{equation}
\label{01}ds^2=(H\sin \eta )^{-2}\left( -d\eta ^2+d\chi ^2+\sin ^2\chi
\left( d\theta ^2+\sin ^2\theta \,d\varphi ^2\right) \right), 
\end{equation}
where $\eta \in (0,\pi )$ and $(\chi ,\theta ,\varphi )$ are the usual
hyperspherical parametrization of the $S^3$ spatial surfaces, which are
invariant under the $O(4)$ subgroup of isometries.

A scalar field of mass $m$ satisfies the Klein-Gordon equation
$\left( \Box -m^2-\xi R\right) \phi (x)=0$, where $\xi$ is a
dimensionless constant which determines the coupling of the field to the
spacetime curvature. We look for a set of mode solutions that can be
written as $u_{klm}(x)=H\sin \eta\,X_k(\eta )\,Y_{klm}(\chi ,\theta
,\varphi)$, where $Y_{klm}$ (with $k=0,1,2,\ldots ;l=0,1,\ldots ,k-1$ and
$m=-l,\ldots ,l$) are the $S^3$ spherical harmonics obeying
$\Delta^{(3)}Y_{klm}=-k(k+2)Y_{klm}$, which constitute a $(k+1)^2$
dimensional representation of $O(4)$. Substituting into the Klein-Gordon
equation, $X_k(\eta )$ satisfy the following equation: 
\begin{equation}
\label{04}\frac{d^2X_k}{d\eta ^2}+\left\{ (k+1)^2+(H\sin \eta )^{-2}\left[
m^2+\left( \xi -\frac 16\right) R\right] \right\} X_k=0. 
\end{equation}
The general solution to this equation is 
\begin{equation}
\label{05}X_k(\eta )=(\sin \eta )^{-\frac 12}\left[ A_kP_{k+\frac 12%
}^\lambda (-\cos \eta )+B_kQ_{k+\frac 12}^\lambda (-\cos \eta )\right], 
\end{equation}
where $\lambda \equiv \sqrt{\frac 94-\frac{12}R(m^2+\xi R)}$, the
coefficients $A_k$ and $B_k$ satisfy the normalization condition
$A_kB_k^{*}-A_k^{*}B_k=i\Gamma (k+\frac 32%
-\lambda )/\Gamma (k+\frac 32+\lambda )$ and $P_{k+\frac 12}^\lambda
(z)$ and $Q_{k+\frac 12}^\lambda (z)$ are associated Legendre functions of
the first
and second kind respectively. Given such a complete set of orthonormal
solutions, one can expand the field operator as $\hat{\phi }%
(x)=\sum_{klm}\left( \hat{a}_{klm}u_{klm}(x)+\hat{a}%
_{klm}^{+}u_{klm}^{*}(x)\right) $ and build the associated Fock space of
states in the usual way. If we require the vacuum state to be $O(4,1)$
invariant and of Hadamard type, $A_k$ and $B_k^{}$ are uniquely determined 
\cite{allen87,chernikov68/schomblond76}: 
\begin{eqnarray}
\label{07}
A_k&=&\frac \pi 4\frac{\Gamma (k+\frac 32-\lambda )}{\Gamma (k+\frac 32
+\lambda)}, \\
B_k&=&-\frac{2i}\pi A_k .
\end{eqnarray}
The corresponding state is the so-called Euclidean vacuum, also known as
Bunch-Davies vacuum.

\section{DISCUSSION ABOUT THE CHOSEN STATE} \label{sec3}

The most natural state to choose is the Euclidean vacuum, which is
invariant
under the de Sitter group of isometries and is a Hadamard state. This is
the
one selected in quantum cosmology when Hartle-Hawking no-boundary proposal
or Vilenkin's tunneling wave function are taken as initial conditions
\cite
{halliwell89}. In addition, when the field is massive, the field state
asymptotically tends to it \cite{hawking83}.

Both the Green's functions and the renormalized expectation value of the
stress-energy tensor have been computed for the Euclidean vacuum in
several
works \cite{bunch78,birrell84}: 
\begin{equation}
\label{08}G^{\left( 1\right) }(x,y)\equiv \left\langle 0\right| \{ 
\hat{\phi }(x),\hat{\phi }(y)\} \left| 0\right\rangle
=\frac{%
2H^2}{(4\pi )^2}\,\Gamma\!\left( \frac 32-\lambda \right) \Gamma\!\left(
\frac 32+\lambda \right) F\!\left( \frac 32-\lambda ,\frac 32+\lambda
,2;\frac{1+Z(x,y)}2\right),
\end{equation}
\begin{eqnarray}
\label{09}
\left\langle 
\hat{T}_{\mu \nu }(x)\right\rangle _{ren}=-\frac{g_{\mu \nu}(x)}{64\pi^2}
\left\{ m^2\left[ m^2+\left( \xi -\frac 16\right) R\right] \left[ \psi\!
\left( \frac 32-\lambda \right) +\psi\! \left( \frac 32+\lambda \right)
+\ln 
\frac R{12m^2}\right] \right. \nonumber \\
\left. -m^2\left( \xi -\frac 16\right) R-%
\frac 1{18}m^2R-\frac 12\left( \xi -\frac 16\right)
^2R^2+\frac{R^2}{2160}%
\right\},
\end{eqnarray}
where $F(a,b,c;z)$ is a hypergeometric function and $\psi (z)\equiv
(d/dz)\,\ln \Gamma (z)$.

From (\ref{08}) one can see that the symmetrized two-point function $%
G^{\left( 1\right) }(x,y)$ has an infrared divergence (as $m^2\rightarrow
0$%
\/) in the minimal coupling case. On the other hand, the expectation value
of the stress-energy tensor remains finite, but is ambiguous, i.e., it
depends on the way the limit $m^2\rightarrow 0$, $\xi \rightarrow 0$ is
taken. In fact, Allen \cite{allen85} proved that in the massless minimally
coupled case there can be no Fock space built on a de Sitter invariant
vacuum. Allen and
Folacci suggested that in this case one should consider instead $O(4)$
invariant states \cite{allen87}. On the other hand Garriga and Kirsten
pointed out that such a peculiar behavior comes from the zero mode, which
corresponds to a constant solution of the Klein-Gordon equation that
appears
when $k=0$ in (\ref{04}) \cite{garriga93}. They also showed that a special
treatment of the zero mode
seems to give an invariant vacuum. The Hilbert space of states is then the
tensor product of the zero mode part, which is equivalent to the Hilbert
space for a one dimensional non-relativistic free particle, times the Fock
space corresponding to the non-zero modes (the whole space is not a Fock
space, in agreement with Allen's result). The field operator is written 
\begin{equation}
\label{001}\hat{\phi }(x)=\sum_{%
{{klm} \atop {k\neq 0}}
}\left( \hat{a}_{klm}u_{klm}(x)+\hat{a}_{klm}^{+}u_{klm}^{*}(x)%
\right) +\frac H{\sqrt{2}\pi }\left( \hat{Q\,}\frac
12+\hat{P}\left(
\eta -\frac 12\sin 2\eta -\frac \pi 2\right) \right) ,
\end{equation}
where $\hat{Q}$ and $\hat{P}$ are the position and momentum
operators of the free particle Hilbert space and $\hat{a}_{klm}^{+}$
and $\hat{a}_{klm}$ are the creation and annihilation operators of the
Fock space associated to all the modes with $k\neq 0$.
Note that the two functions that multiply the operators $\hat{Q}$ and
$\hat{P}$, when divided by $\sin \eta$, are solutions of equation
(\ref{04}) for $k=0$ (with $m=\xi=0$). In fact, the constant
function multiplying the operator $\hat{Q}$ is precisely the aforesaid
zero mode. These operators satisfy the usual commutation relations: 
\begin{equation}
\label{002}
\begin{array}{c}
\left[ \hat{a}_{klm},\hat{a}_{k^{\prime }l^{\prime }m^{\prime }}\right]
=0,\qquad \left[ \hat{a}_{klm},\hat{a}_{k^{\prime }l^{\prime}
m^{\prime }}^{+}\right] =\delta _{kk^{\prime }}\delta _{ll^{\prime}}
\delta_{mm^{\prime }},\qquad [ \hat{Q},\hat{P} ] =i, \\ 
\protect{} [ \hat{a}_{klm},\hat{Q} ] =0,\qquad [ \hat{a}_{klm},\hat{P} ]
=0 .
\end{array}
\end{equation}
The vacuum is defined as follows: 
\begin{equation}
\label{003}
\begin{array}{c}
\hat{P}\left| 0\right\rangle =0 ,\\ \hat{a}_{klm}\left|
0\right\rangle =0\;,\,k\neq 0 .
\end{array}
\end{equation}
As a consequence of the first condition, this vacuum is not normalizable,
so strictly speaking it is not an
element of the Hilbert space but just a limit. However, as it has been
pointed out by Garriga and Kirsten, this fact is not so
strange: it also happens for the ground state of a non-relativistic free
particle. These authors also computed the renormalized expectation
value of the stress-energy tensor for this state: 
\begin{equation}
\label{2}\left\langle \hat{T}_{\mu \nu }(x)\right\rangle
_{ren}=\frac{119}{138240\pi ^2}\,R^2\,g_{\mu \nu }(x)=\frac{119}{960\pi
^2}\,H^4\,g_{\mu \nu }(x). 
\end{equation}
Two aspects should be stressed. First, the expression is de Sitter
invariant. Second, the energy density is lower than that of the $O(4)$
invariant
states introduced by Allen an Folacci, the stress-energy tensor of which
is not de Sitter invariant. It is also remarkable that the expectation
value for other states of observables such as the dispersion of the
smeared field or the stress-energy tensor tend
asymptotically ($\eta \rightarrow \pi $ or equivalently $t\rightarrow
\infty $, where $t$ is the proper time for a comoving observer) to the
same value as that of the invariant vacuum \cite{garriga93}.

\section{COMPUTATION OF THE STRESS-ENERGY TENSOR FLUCTUATIONS}
\label{sec4}

The quantity that characterizes the stress tensor fluctuations and which
determines, via the Einstein-Langevin equation, the fluctuations in the
spacetime geometry \cite{martin98} is the noise kernel, defined by the
symmetrized two-point correlation function $\frac 12\left\langle
\{\hat{t}_{\mu \nu}(x),\hat{t}_{\rho \sigma }(y)\} \right\rangle = \Re
\left\langle
\hat{t}_{\mu \nu}(x)\hat{t}_{\rho \sigma }(y) \right\rangle$, 
where $\hat{t}_{\mu \nu}
(x)\equiv \hat{T}_{\mu \nu }(x)-\left\langle \hat{T}_{\mu \nu
}(x)\right\rangle $ and $\hat{T}_{\mu \nu }(x)\equiv\lim_{x^{\prime
}\rightarrow x}{\cal D}_{\mu \nu}\,(x,x^{\prime})\left( \hat{\phi
}(x)\hat{\phi
}(x^{\prime })\right) $ with ${\cal D}_{\mu \nu }(x,x^{\prime })\equiv
\left(\nabla_\mu ^x\nabla _\nu ^{x^{\prime}}-\frac 12g_{\mu \nu }(x)\nabla
_\alpha ^x\nabla ^{x^{\prime }\alpha }\right)$ (we have chosen
point-splitting regularization for convenience). It is easy to see that
$\left\langle
\hat{t}_{\mu \nu}(x)\hat{t}_{\rho \sigma }(y) \right\rangle$
is equivalent to $\left\langle \hat{T}_{\mu \nu }(x)\hat{T}_{\rho
\sigma }(y)\right\rangle -\left\langle \hat{T}_{\mu \nu
}(x)\right\rangle \left\langle \hat{T}_{\rho \sigma }(y)\right\rangle$.
Note that this expression is finite in the following sense: one can
compute it suitably regularized, then the potentially divergent terms
cancel and one can remove the regularization (letting $x^{\prime
}\rightarrow x$ and $y^{\prime }\rightarrow y$). 
\begin{eqnarray}
\label{3}
\left\langle 
\hat{T}_{\mu \nu }(x)\hat{T}_{\rho \sigma }(y)\right\rangle
-\left\langle \hat{T}_{\mu \nu }(x)\right\rangle \left\langle
\hat{T}_{\rho \sigma }(y)\right\rangle =\lim_{{x^{\prime }\rightarrow
x}\atop{y^{\prime }\rightarrow y}}{\cal D}_{\mu \nu }(x,x^{\prime}){\cal
D}_{\rho\sigma}(y,y^{\prime})&&\left( \left\langle \hat{\phi}(x)\hat{\phi}
(x^{\prime })\hat{\phi }(y)\hat{\phi }(y^{\prime })\right\rangle \right. 
\nonumber \\ 
&&\left. -\left\langle \hat{\phi }(x)\hat{\phi }(x^{\prime
})\right\rangle \left\langle \hat{\phi }(y)\hat{\phi }(y^{\prime
})\right\rangle \right). 
\end{eqnarray}
The expression inside the parenthesis is computed in the appendix, where
we find that it is finite as we pointed out above, and equals
$G^{+}(x,y)G^{+}(x^{\prime},y^{\prime})+G^{+}(x,y^{\prime})G^{+}
(x^{\prime},y)$. The final result after removing the regularization is: 
\begin{eqnarray}
\label{6}
\left\langle 
\hat{t}_{\mu \nu }(x)\hat{t}_{\rho \sigma }(y)\right\rangle &=&
\Bigl[
 \nabla _\rho ^y\nabla _\mu ^xG^{+}(x,y)\nabla _\sigma ^y\nabla _\nu
^xG^{+}(x,y)+\nabla _\sigma ^y\nabla _\mu ^xG^{+}(x,y)\nabla _\rho
^y\nabla
_\nu ^xG^{+}(x,y) \nonumber \\ 
&& -g_{\mu \nu }(x)\nabla _\rho
^y\nabla _\alpha ^xG^{+}(x,y)\nabla _\sigma ^y\nabla ^{x\alpha
}G^{+}(x,y)-g_{\rho \sigma }(y)\nabla _\alpha ^y\nabla _\mu
^xG^{+}(x,y)\nabla ^{y\alpha }\nabla _\nu ^xG^{+}(x,y) \nonumber \\ 
&& +\frac 12g_{\mu \nu }(x)g_{\rho \sigma }(y)\nabla _\beta ^y\nabla
_\alpha ^xG^{+}(x,y)\nabla ^{y\beta }\nabla ^{x\alpha }G^{+}(x,y)\Bigr]. 
\end{eqnarray}

Thus, we need to determine the Wightman function $G^{+}(x,y)\equiv
\left\langle 0\right| \hat{\phi }(x) \hat{\phi }(y)\left|
0\right\rangle $ for our chosen state. If we consider the Euclidean vacuum
for a massive minimally coupled field and take the massless limit, the
Wightman function diverges. As already explained, this is connected to
the impossibility of having a Fock space built on a de Sitter invariant
vacuum in the massless case. To apply Garriga and Kirsten's special
treatment of the zero mode, we have to consider separately the
contribution
from the zero mode and that from all the rest. To compute the latter, one
can use the expression for the massive case, with the mass acting as a
regulator of the infrared divergence, subtract the contribution from the
zero mode, which contains all the infrared divergences, and then remove
the regulator, i.e., take the massless limit.

When $x$ and $y$ are spacelike separated, $G^{+}(x,y)=\frac 12\left(
G^{\left( 1\right) }(x,y)+G(x,y)\right) =\frac 12 G^{\left( 1\right)
}(x,y)= \Re G^{+}(x,y)$
since the commutator $G(x,y)\equiv \left\langle 0\right| [ \hat{%
\phi }(x),\hat{\phi }(y)] \left| 0\right\rangle $ vanishes for
causally disconnected points. In this case the noise kernel coincides with
expression (\ref{6}). We can also take advantage of the results in refs.
\cite{garriga93,allen87} to write: 
\begin{equation}
\label{7}G^{\left( 1\right) }(x,y)=G_{NZM}^{\left( 1\right)
}(x,y)+\frac{H^2%
}{8\pi ^2}\left\langle 0\right| \hat{Q}^2\left| 0\right\rangle ,
\end{equation}
where
\begin{equation}
\label{8}G_{NZM}^{\left( 1\right) }(x,y)\equiv\frac R{48\pi ^2}\left[
\frac 1{1-Z(x,y)}-\ln \left( 1-Z(x,y)\right) -\ln (4\sin \eta _x\sin \eta
_y)-\sin^2\eta _x-\sin ^2\eta _y\right] .
\end{equation}
Several remarks are in order. $G_{NZM}^{\left( 1\right) }(x,y)$
corresponds
to the non-zero mode contribution whereas the contribution from the zero
mode reduces to $H^2/8\pi ^2 \,\left\langle 0\right| \hat{Q}%
^2\left| 0\right\rangle $ since $\hat{P}\left| 0\right\rangle =0$.
This
term is actually divergent but is independent of $x$ and $y$, so that it
will give no contribution to (\ref{6}). Furthermore, the term $\ln
(2\sin \eta _x)+\ln (2\sin \eta _y)+\sin ^2\eta _x+\sin ^2\eta _y$
will not contribute either, because in (\ref{6}) there are always
derivatives with respect to both $x$ and $y$ acting on $G^{+}(x,y)$.
Consequently, we only need to take into account the following part of the
two-point function $G^{(1)}(x,y)$:
\begin{equation}
\label{9}{\cal G}(x,y)\equiv \frac R{48\pi ^2}\left[ \frac 1{1-Z(x,y)}-\ln
\left( 1-Z(x,y)\right) \right] .
\end{equation}
If we consider spacelike separated points which are geodesically
connected,
we can use expression (\ref{0}) to derive the following results: 
\begin{eqnarray}
\label{10}
\nabla _\mu ^x\sigma (x,y)=s(x,y)s_\mu (x), \\
\label{11}
\nabla _\rho ^y\nabla _\mu ^x\sigma (x,y)=s_\rho (y)s_\mu (x),
\end{eqnarray}
where $s^\mu (x)$ is the unit vector tangent at point $x$ to the geodesic
joining $x$ and $y$. The derivatives of $Z(x,y)$ are then 
\begin{eqnarray}
\label{12}
\nabla _\mu ^xZ(x,y)=
\sqrt{\frac R{12}\left( 1-Z^2(x,y)\right) }\;s_\mu (x), \\ \nabla _\rho
^y\nabla _\mu ^xZ(x,y)=\frac R{12}\left( 1-Z^2(x,y)\right) s_\rho (y)
s_\mu (x).
\end{eqnarray}
Thus, after a few algebraic manipulations, we have 
\begin{equation}
\label{13}
\nabla _\rho ^y\nabla _\mu ^xG^{+}(x,y)=
\nabla _\rho ^y\nabla _\mu ^x{\cal G}(x,y)=\frac{H^4}{4\pi ^2}
\frac{1+4Z(x,y)}{\left( 1-Z(x,y)\right) ^2}\,s_\rho (y)s_\mu (x) .
\end{equation}
Substituting this result into (\ref{6}), we get our final result for the
noise kernel corresponding to spacelike separated points: 
\begin{eqnarray}
\label{14}
\left\langle 
\hat{t}_{\mu \nu }(x)\hat{t}_{\rho \sigma }(y)\right\rangle
=\frac{H^8}{16\pi ^4}\left( \frac{1+4Z(x,y)}{\left( 1-Z(x,y)\right)
^2}\right)
^2 && \bigl[ 2 s_\mu (x)s_\nu (x)\,s_\rho (y)\,s_\sigma (y)+g_{\mu \nu
}(x)\,s_\rho (y)\,s_\sigma (y) \nonumber \\
&&  +g_{\rho \sigma }(y)s_\mu
(x)s_\nu (x)+\frac 12g_{\mu \nu }(x)g_{\rho \sigma }(y)\bigr] .
\end{eqnarray}

\section{CONCLUSION}

Comparing (\ref{14}) with the ``square'' of (\ref{2}) one realizes that
the
contribution from the stress-energy fluctuations is at least as important
as that coming from the expectation value. Note that in order to compare
both
expressions it is convenient to consider the metric and the tangent vector
components as referred to an orthonormal base. 
Of course, given that $H^4 << m_p^2 H^2$, this stochastic back-reaction
source is still much smaller than the dominant term which
drives inflation (usually coming from the potential of the inflaton scalar
field) and which is responsible for the near de Sitter geometry of the
spacetime. It seems interesting to point out that the
expectation value (\ref{2}) yields a negative value for the energy density
because of two facts. First, Ford and collaborators (see for instance ref.
\cite{kuo93}) have suggested that one may expect important stress-energy
fluctuations especially in those cases where the energy density is
negative. Second, this is in agreement with some of the results found in
\cite{abramo97b}.

It is only the contribution to the variations of the potential coming from
fluctuations of the inflaton field which are usually considered when
addressing the generation of large-scale gravitational inhomogeneities.
Expressions of the sort $\left\langle \phi ^2(x)\right\rangle
=\frac{H^3}{4\pi ^2}\Delta t$ or related ones \cite{ford82/linde82},
where proper infrared (related to initial conditions) and ultraviolet
cut-offs (only scales larger than the horizon are considered:
$\lambda_{phys}>H^{-1}$) have been imposed, are of frequent use in the
literature \cite{linde90,starobinsky82}.
These fluctuations for the smeared inflaton field can be interpreted as if
the smeared field were undergoing a sort of ``Brownian motion''
\cite{vilenkin83}. Such an
interpretation still applies to Garriga and Kirsten's vacuum as they show
when analyzing the dispersion of the smeared field \cite{garriga93}.
The fluctuations associated to a given cosmic scale arise from the
stochastic contribution, when leaving the horizon, from the modes of the
inflaton field perturbations with a wavelength corresponding to that
scale. The usual treatment just deals with first order variations of the
potential, which are assumed to provide the dominant contribution, and
does not take into account the back-reaction contribution from the kinetic
terms of the quantum perturbations of the inflaton field \cite{linde90}.
The latter can be modeled, at least partially, by a free massless
minimally coupled quantum scalar field. This is precisely the case we
considered in this paper. Furthermore, as mentioned above, we have found
that the contribution from the fluctuations of the stress-energy tensor
can be as important as the expectation value itself. Within the
inflationary context, this contribution to back reaction becomes
especially relevant above the ``self-reproduction'' scale and could have
important implications for the stochastic inflation approach, where this
effect is never taken into account.

One can try to extract information about large-scale fluctuations from
correlations
for spacelike separated points which are beyond the horizon distance \cite
{hawking82}. Unfortunately, for such points $Z(x,y)<-1$ and there is no
geodesic connecting them, thus one cannot directly use our results, which
were obtained using (\ref{10}) and (\ref{11}), which in turn rely on
the particular expression (\ref{0}) for geodesically connected points. If
one is interested in the $Z(x,y)<-1$ case, then the general expression
(\ref{00}) should be used. The relative importance of the fluctuations
that we have found for geodesically connected points encourages us to
pursue this research further.

\section*{ACKNOWLEDGMENTS}

It is a pleasure to thank Esteban Calzetta, Larry Ford, Jaume Garriga,
Bei-Lok Hu, Rosario Mart\'\i n and Xavier Montes for useful comments and
stimulating discussions. This work has been partially supported by the
CICYT
Research Project number AEN95-0590, and the European Project number
CI1-CT94-0004. A.R. also acknowledges support of a grant from the
Generalitat de Catalunya.

\appendix
\section*{}

In this appendix we will compute $\langle \hat{\phi}(x)\hat{%
\phi }(x^{\prime })\hat{\phi }(y)\hat{\phi }(y^{\prime
})\rangle -\langle \hat{\phi }(x)\hat{\phi }(x^{\prime
})\rangle \langle \hat{\phi }(y)\hat{\phi }(y^{\prime
})\rangle $ in the context of Garriga and Kirsten's treatment. So we
will use expression (\ref{001}) for the field $\hat{\phi }(x)$ and
take the expectation values with respect to the vacuum state defined in
(\ref{003}). First, we simply rewrite (\ref{001}) for convenience: 
\begin{equation}
\label{60}\hat{\phi }(x)=\sum_{%
{{klm} \atop {k\neq 0}}
}\left( \hat{a}_{klm}u_{klm}(x)+\hat{a}_{klm}^{+}u_{klm}^{*}(x)%
\right) +\left( \hat{Q}\,\frac{u_{000}(x)}{\left\langle 0\right|
\hat{Q}^2\left| 0\right\rangle^\frac 12}+\hat{P}\,v_0(x)\right) .
\end{equation}
Using the commutation relations (\ref{002}), we find 
\begin{eqnarray}
\label{61}
\left\langle 
\hat{\phi }(x)\hat{\phi }(x^{\prime })\hat{\phi
}(y)\hat{\phi }(y^{\prime })\right\rangle &=&
\sum_{klm}\sum_{k^{\prime }l^{\prime
}m^{\prime }}\left( u_{klm}(x)u_{klm}^{*}(y)u_{k^{\prime }l^{\prime
}m^{\prime }}(x^{\prime })u_{k^{\prime }l^{\prime }m^{\prime
}}^{*}(y^{\prime })+u_{klm}(x)u_{klm}^{*}(y^{\prime })u_{k^{\prime
}l^{\prime }m^{\prime }}(x^{\prime })u_{k^{\prime }l^{\prime }m^{\prime
}}^{*}(y)\right) \\ && + \sum_{klm}u_{klm}(x)u_{klm}^{*}(x^{\prime
})\sum_{k^{\prime }l^{\prime }m^{\prime }}u_{k^{\prime }l^{\prime
}m^{\prime
}}(y)u_{k^{\prime }l^{\prime }m^{\prime }}^{*}(y^{\prime }) ,
\end{eqnarray}
and 
\begin{equation}
\label{62}\left\langle \hat{\phi }(x)\hat{\phi }(x^{\prime
})\right\rangle \left\langle \hat{\phi }(y)\hat{\phi }(y^{\prime
})\right\rangle =\sum_{klm}u_{klm}(x)u_{klm}^{*}(x^{\prime
})\sum_{k^{\prime
}l^{\prime }m^{\prime }}u_{k^{\prime }l^{\prime }m^{\prime
}}(y)u_{k^{\prime
}l^{\prime }m^{\prime }}^{*}(y^{\prime }) .
\end{equation}
From this we obtain
\begin{equation}
\label{63}\left\langle \hat{\phi }(x)\hat{\phi }(x^{\prime })%
\hat{\phi }(y)\hat{\phi }(y^{\prime })\right\rangle -\left\langle 
\hat{\phi }(x)\hat{\phi }(x^{\prime })\right\rangle \left\langle 
\hat{\phi }(y)\hat{\phi }(y^{\prime })\right\rangle
=G^{+}(x,y)G^{+}(x^{\prime },y^{\prime })+G^{+}(x,y^{\prime
})G^{+}(x^{\prime },y) ,
\end{equation}
where $G^{+}(x,y)$ is the Wightman two-point function
\begin{equation}
\label{64}G^{+}(x,y)=\sum_{%
{{klm} \atop {k\neq 0}}
}u_{klm}(x)u_{klm}^{*}(y)+\,u_{000}(x)\,u_{000}^{*}(x)\,\left\langle
0\right| \hat{Q}^2\left| 0\right\rangle .
\end{equation}


\end{document}